# Frequency-domain "single-shot" (FDSS) ultrafast transient absorption spectroscopy using compressed laser pulses. Part II. Supplemetary Material. [1]

Ilya A. Shkrob[*,a], Dmitri A. Oulianov,[a] Robert A. Crowell,[a] and Stanislas Pommeret [a,b]

[a] Chemistry Division , Argonne National Laboratory, Argonne, IL 60439
[b] CEA/Saclay, DSM/DRECAM/SCM/URA 331 CNRS 91191 Gif-Sur-Yvette Cedex, France

**Abstract**

Single-shot ultrafast absorbance spectroscopy based on the frequency encoding of the kinetics is analyzed theoretically and implemented experimentally. In Part II of the series, arbitrary thickness sample is analysed therotetically. The model is then used to simulate the results for a-Si:H films.

## Appendix 1. Arbitrary thickness sample on a clear substrate.

In this Appendix, we examine a general case of the probe light propagating through a thin sample. Without loss of generality, we will assume that the sample is a flat layer of thickness $d$ with a complex dielectric function $\varepsilon_\omega = n_\omega^2$ on a thick, non-absorbing substrate with index of refraction $n_s$. The propagation of light through the sample in the direction normal to the surface is given by a wave equation

$$c^2 \frac{\partial^2 E(t,z)}{\partial z^2} = \{\varepsilon + \Delta\varepsilon(t)\} \otimes \frac{\partial^2 E(t,z)}{\partial t^2} \tag{A1}$$

where $z$ is the depth of the sample ($z=0$ for the air/sample interface and $z=d$ for the sample/substrate interface) and $\Delta\varepsilon(t)$ is the perturbation of the dielectric function by the pump light. We will assume that this perturbation is independent of $z$, i.e., the pump light is absorbed homogeneously throughout the sample (if the energy deposition is $z$-dependent, even the PPS experiment becomes difficult to analyze rigorously, see refs. 24 and 1S). To solve eq. (A1) we will use first order perturbation theory and assume that

$$E(t,z) \approx E^0(t,z) + \Delta E(t,z) \tag{A2}$$

where the zero and the first order terms satisfy the equations

$$c^2 \frac{\partial^2 E^0(t,z)}{\partial z^2} = \varepsilon \frac{\partial^2 E^0(t,z)}{\partial t^2} \tag{A3}$$

$$c^2 \frac{\partial^2 \Delta E(t,z)}{\partial z^2} - \varepsilon \frac{\partial^2 \Delta E(t,z)}{\partial t^2} = \Delta\varepsilon(t) \otimes \frac{\partial^2 E^0(t,z)}{\partial t^2} \qquad (A4)$$

The first of these equations has the solution

$$E^0(t,z) = \int_{-\infty}^{\infty} d\omega \; e^{-i\omega t}\left(E_\omega^{0,+} e^{+ik_\omega z} + E_\omega^{0,-} e^{-ik_\omega z}\right) \qquad (A5)$$

where $k_\omega = \omega n_\omega / c$ is the wave vector of the light with frequency $\omega$ and $n_\omega = \eta_\omega + i\kappa_\omega$ is the complex refraction index. Coefficients $E_\omega^{0,\pm}$ in eq. (A5) may be found from the continuity of the electric field $E(t,z)$ and its derivative $\partial E(t,z)/\partial z$ with respect to the co-ordinate $z$ at $z=0$ and $z=d$:

$$
\begin{aligned}
(1+r_\omega)E_\omega^i &= E_\omega^{0,+} + E_\omega^{0,-} \\
(1-r_\omega) E_\omega^i &= n_\omega (E_\omega^{0,+} - E_\omega^{0,-}) \\
t_\omega E_\omega^i &= E_\omega^{0,+} e^{i\delta_\omega} + E_\omega^{0,-} e^{-i\delta_\omega} \\
n_s \, t_\omega E_\omega^i &= n_\omega (E_\omega^{0,+} e^{i\delta_\omega} - E_\omega^{0,-} e^{-i\delta_\omega})
\end{aligned}
\qquad (A6)
$$

where $r_\omega$ and $t_\omega$ are the reflection and transmission (Fresnel) coefficients, $\delta_\omega = k_\omega d$ is the complex phase acquired by the Fourier component with the frequency $\omega$ as it propagates through the sample, and $E_\omega^i$ is the amplitude of the incident light. Eqs. (A6) can be solved to obtain the Fresnel coefficients: [1S]

$$t_\omega = 2n_\omega / D_\omega \quad and \quad r_\omega = \{\cos\delta_\omega \; n_\omega (1-n_s) + i \sin\delta_\omega \; (n_\omega^2 - n_s)\}/D_\omega \qquad (A7)$$

where

$$D_\omega = \cos\delta_\omega \; n_\omega (1+n_s) - i \sin\delta_\omega \; (n_\omega^2 + n_s) \qquad (A8)$$

Neglecting the reflections from the back surface of the substrate, the transmission coefficient for the light passing through the thin sample on the thick substrate is equal to $t_s t_\omega$, where $t_s = 2n_s/(n_s+1)$ is the transmission coefficient for the substrate. The photoinduced change $\Delta t_\omega$ in the Fresnel coefficient causes a change in the light transmission:

$$S(\omega) = (|t_\omega|^2 - |t_\omega + \Delta t_\omega|^2)/|t_\omega|^2 \approx -2\,\text{Re}\,(\Delta t_\omega / t_\omega) \qquad (A9)$$

To estimate $\Delta t_\omega$, we rewrite the right side of eq. (A4) as

$$\Delta\varepsilon(t) \otimes \frac{\partial^2 E^0(t,z)}{\partial t^2} = -\int_{-\infty}^{\infty} d\Omega \; \Omega^2 \; \Delta\varepsilon(t) \otimes e^{-i\Omega t}\left(E_\Omega^{0,+} e^{+ik_\Omega z} + E_\Omega^{0,-} e^{-ik_\Omega z}\right) \qquad (A10)$$

The Fourier component $\Delta E_\omega(z)$ of $\Delta E(t,z)$ is given by

$$c^2 \frac{\partial^2 \Delta E_\omega(z)}{\partial z^2} + \varepsilon_\omega \omega^2 \Delta E_\omega(z) = -\int_{-\infty}^{\infty} d\Omega \; \Omega^2 \; K_{\Omega-\omega}\left(E_\Omega^{0,+} e^{+ik_\Omega z} + E_\Omega^{0,-} e^{-ik_\Omega z}\right) \qquad (A11)$$

where the function $K_{\Omega-\omega}$ is given by eq. (9). The general solution of this equation is given by

$$\Delta E_\omega(z) = \int_{-\infty}^{\infty} d\Omega \; \frac{\Omega^2 \; K_{\Omega-\omega}}{\Omega^2 \varepsilon_\Omega - \omega^2 \varepsilon_\omega} \; \Delta E_{\omega,\Omega}(z) \qquad (A12)$$

where

$$\Delta E_{\omega,\Omega}(z) = E_\Omega^{0,+} e^{+ik_\Omega z} + E_\Omega^{0,-} e^{-ik_\Omega z} + C_\Omega^+ e^{+ik_\omega z} + C_\Omega^- e^{-ik_\omega z} \qquad (A13)$$

and the coefficients $C_\Omega^\pm$ satisfy the boundary conditions:

$$\Delta r_{\omega,\Omega} E^i_\Omega = \Delta E_{\omega,\Omega}(z=0)$$

$$-\Delta r_{\omega,\Omega} E^i_\Omega \; \omega/c = \Delta E'_{\omega,\Omega}(z=0)$$

$$\Delta t_{\omega,\Omega} E^i_\Omega = \Delta E_{\omega,\Omega}(z=d) \tag{A14}$$

$$\Delta t_{\omega,\Omega} E^i_\Omega \; \omega n_s/c = \Delta E'_{\omega,\Omega}(z=d)$$

where $\Delta r_{\omega,\Omega}$ and $\Delta t_{\omega,\Omega}$ are the corrections to the corresponding Fresnel coefficients. The overall change $\Delta t_\omega$ in the transmission coefficient is given by

$$\Delta t_\omega E^i_\omega = \int_{-\infty}^{+\infty} d\Omega \; \frac{\Omega^2 \; K_{\Omega-\omega}}{\Omega^2 \varepsilon_\Omega - \omega^2 \varepsilon_\omega} \; \Delta t_{\omega,\Omega} \; E^i_\Omega \tag{A15}$$

Eqs. (A14) may be combined with eqs. (A6) to obtain, after some algebra,

$$\Delta t_{\omega,\Omega} = t_\Omega - t_\omega + \frac{k_\Omega - k_\omega}{2 k_\omega} t_\omega \left\{ r_\Omega - 1 + n_s t_\Omega \left( \cos\delta_\omega - \frac{i}{n_\omega} \sin\delta_\omega \right) \right\} \tag{A16}$$

Assuming that $\Delta\varepsilon_\omega(t)$ has a time-independent spectral profile and $\Omega$ is sufficiently close to $\omega$, we can approximate

$$\frac{\Omega^2 \; K_{\Omega-\omega}}{\Omega^2 \varepsilon_\Omega - \omega^2 \varepsilon_\omega} \approx \frac{\Delta k_\omega}{k_\Omega - k_\omega} \; \hat{K}_{\Omega-\omega} \tag{A17}$$

to obtain the final result:

$$S(\omega) = -2 \; \text{Re} \left\{ \Delta k_\omega \int_{-\infty}^{+\infty} d\Omega \; \hat{K}_{\Omega-\omega} \frac{E^i_\Omega}{E^i_\omega} \Theta_\omega(\Omega) \right\} \tag{A18}$$

where

$$\Theta_\omega(\Omega) = \frac{t_\Omega/t_\omega - 1}{k_\Omega - k_\omega} + \frac{1}{2 k_\omega} \left\{ r_\Omega - 1 + n_s t_\Omega \left( \cos\delta_\omega - \frac{i}{n_\omega} \sin\delta_\omega \right) \right\} \tag{A19}$$

Formula (A18) resembles eq. (12) obtained in section 2. In particular, it is easy to demonstrate, by direct calculation, that

$$\Delta k_\omega \; \Theta_\omega(\omega) = \frac{1}{t_\omega} \left( \frac{\partial t}{\partial \omega} \bigg/ \frac{dk}{d\omega} + \frac{\partial t}{\partial k} \right)_\omega \Delta k_\omega = \frac{1}{t_\omega} \left( \frac{dt}{d\varepsilon} \right)_\omega \Delta\varepsilon_\omega \tag{A20}$$

Thus, if the sample is sufficiently thin, so that $n_\omega d \ll c\tau_p$ (i.e., when $\Theta(\Omega)$ is a relatively slow function of $\Omega$ over the spectral range of interest), we obtain

$$S(\omega) \approx -2 \; \text{Re} \left\{ \frac{1}{t_\omega} \left( \frac{dt}{d\varepsilon} \right)_\omega \Delta\varepsilon_\omega \int_{-\infty}^{+\infty} d\Omega \; \hat{K}_{\Omega-\omega} \frac{E^i_\Omega}{E^i_\omega} \right\} \tag{A21}$$

For single exponential kinetics, the integral in eq. (A21) is equal to the function $\Phi(\alpha,\beta,\gamma)$ given by eq. (14). In particular, for long group delays $T_e$, the integral in eq. (A21) asymptotically approaches $exp(-\gamma T_e)$ (this applies to the samples of any thickness, since for long $T_e$, a very narrow interval of frequencies $\Omega$ close to $\omega$ contributes to the integral in eq. (A18)). For a very thin sample $(\delta_\omega \ll 1)$,

$$t_\omega \approx \frac{2}{n_s + 1} \quad \text{and} \quad \frac{1}{t_\omega} \left( \frac{dt}{d\varepsilon} \right)_\omega \approx i t_\omega \omega d / 2c \tag{A22}$$

so that for $n_s = 1$ (no substrate) we obtain eq. (12) recast as

$$S(\omega) = \frac{\omega d}{c} \operatorname{Im} \left\{ \Delta\varepsilon_\omega \int_{-\infty}^{+\infty} d\Omega \ \hat{K}_{\Omega-\omega} \frac{E_\Omega^i}{E_\omega^i} \right\} \tag{A23}$$

(Note that for a thin wedge this expression should be divided by $n_\omega$). In the general case, formula (A21) is incorrect and eqs. (A18) and (A19) must be used instead. The integration can be preformed numerically or by expansion of the Fresnel coefficients given by eq. (A7) (that are periodic functions of $\omega$) into a truncated Fourier series and integrating each term analytically, using a modified eq. (14). In most experimental situations, the second term in eq. (A19) is 1-2 orders of magnitude smaller than the first term and thereby may be neglected. Note that the expression for $S(\omega)$ given by eq. (A18) is for *infinite* spectral resolution. For $\delta\tau_p \ll 1$ (which is always the case experimentally), $S(\omega) \approx S(\omega)|_{\delta=0} \otimes g(\omega)$. The latter convolution is carried out numerically.

We turn now to the thin-film a-Si:H sample examined in section 5.2. From the existing data on the refraction and absorption in a-Si:H, we may safely assume that $n_\omega$ and $\Delta n_\omega$ are frequency independent in a narrow band around 12500 cm$^{-1}$ and let $\eta=3.44$ and $\kappa=1.1 \times 10^{-3}$ be constant.[24] Fig. 1S shows the function

$$\mathcal{T}'(\omega) \approx -2 \operatorname{Re} \left\{ \frac{1}{t_\omega} \left( \frac{dt}{d\varepsilon} \right)_\omega \exp(i\phi_\varepsilon) \right\} \tag{A24}$$

calculated using Fresnel coefficients given by eqs. (A7) and (A8) for $d=1.27$ μm and $\phi_\varepsilon=0^0$ and $90^0$. For $\gamma T_e \gg 1$, $S(\omega)/\mathcal{T}'(\omega) \approx |\Delta\varepsilon_\omega| \exp(-\gamma T_e)$. For pure photoabsorption ($\phi_\varepsilon=90^0$), function $\mathcal{T}'(\omega)$ has the same extrema as the sample transmission $T_\omega = |t_\omega|^2$. For pure photorefraction ($\phi_\varepsilon=0^0$), $\mathcal{T}'(\omega)$ has zeroes at the transmission extrema and maxima and minima at the corresponding inflection points. For the parameters given above, $\mathcal{T}'(\omega)$ has a maximum exactly at the center frequency $\omega_0$ of the probe pulse, and the fringe period is close to the FWHM of this pulse. The calculation indicates that the phase of the complex factor $t_\omega^{-1}(\partial t_\omega/\partial\varepsilon)$ in eqs. (A21) and (A24) changes from $90^\circ$ at the spectral center (where the transmission is minimum) to $\pm 65^\circ$ at the limits of the optimum spectral range (where the reflection has extrema). Cursory examination of eq. (A21) and Fig. 5 suggests that the oscillation pattern of $S(\omega)$ should change with the pump delay $T$ since the kinetic origin $\Delta\omega(T_e = 0) \equiv T/\phi''(\omega)$ moves relative to the fringe pattern and this changes the phase of the complex factor $t_\omega^{-1}(\partial t_\omega/\partial\varepsilon)$. Experimentally, the oscillation pattern barely changes when the kinetic origin is swept across the spectrum. We argue that eq. (A21) is incorrect when the fringe spacing is comparable to $1/\tau_p$. General eqs. (A18) and (A19) must be used in such a case:

Fig. 1S exhibits FDSS kinetics obtained for $T=0$ (kinetic origin at the transmission minimum), $T= 35$ ps (kinetic origin at the reflection maximum), and $T=20$ ps for a photoabsorption signal ($\phi_\varepsilon=90$) that exponentially decays with $\gamma^{-1}=30$ ps. Although these kinetics change considerably as a function of $T$, most of this change is in the weighting factor $t_\omega^{-1}(\partial t_\omega / \partial \varepsilon)$. In Fig. 2S(a), normalized signals $S(\omega)/\mathcal{T}'(\omega)$ are plotted; these normalized kinetics hardly change between these three delays. In Fig. 2S(b), $S(\omega)/\mathcal{T}'(\omega)$ kinetics are plotted as a function of $T_e$ shifted by $T$, in order to juxtapose their oscillation patterns. It is seen that the changes in the positions of crests are small, ca. 20-30% of what would be expected from eq. (A21). E. g., for $T$=20 ps and 35 ps the spacing between the first pair of crests is just 4.5% and 9.1%, respectively, lower than the same spacing for $T$=0 ps. This result is in full agreement with the experimental observations of Appendix 2.

## Appendix 2. FDSS of thin-film a-Si:H alloy.

A PPS measurement of $\Delta\varepsilon_\omega(t)$ becomes quite complicated when the sample exhibits well-resolved interference fringes near the probe wavelength. [24,31,1S] Such a situation frequently occurs in studies of thin films. [1S] In this case, the complex factor $t_\omega^{-1}(\partial t_\omega/\partial\varepsilon)$ in eq. (7) oscillates with $\omega$, and in order to extract $\Delta\varepsilon_\omega(t)$ one needs to (i) determine this factor in a separate experiment, (ii) measure transient reflection of the sample in addition to the transmission, and (iii) invert linear equations that express these two signals vs. $\Delta\varepsilon'_\omega(t)$ and $\Delta\varepsilon''_\omega(t)$. [1S] Involved as it is, this procedure (known as the "inversion method") [1S] is insufficient to obtain $\Delta\varepsilon'_\omega(t)$ near the fringe extrema, where $\mathrm{Im}\, dt_\omega/d\varepsilon = 0$. On the other hand, precisely due to the latter equation, $\Delta\varepsilon''_\omega(t)$ can be determined from a single transmission measurement near the fringe extrema, even if $\Delta\varepsilon'_\omega(t)$ is large, provided that the fringe spacing is much greater than the spectral width of the probe pulse, $1/\tau_p$. [24,1S] The retrieval of $\Delta\varepsilon_\omega(t)$ is further complicated by inhomogeneous absorption of the pump that changes the inversion matrix. Moon and Tauc [1S] found that this method completely breaks down for large photoinduced signals (>0.01). Even when the signals are weak, the results are very sensitive to small errors in $t_\omega^{-1}(\partial t_\omega/\partial\varepsilon)$, [1S] which is seldom known with the required accuracy for the spot probed with the laser light. As shown below, FDSS might be, in some ways, preferable to the inversion method for thin-film samples. The advantage of FDSS is that it inherently combines spectral and kinetic measurement in a single experiment whereas the inversion relies on several independent measurements.

For this demonstration we have chosen amorphous hydrogenated silicon (*a*-Si:H) which is a commercial thin-film material for solar energy conversion. [31] This material has an optical gap 1.75 eV and is transparent at our probe wavelength, 800 nm. The dynamics of photoinduced free carriers and trapped charges in *a*-Si:H has been extensively studied (see reviews in refs. 24, 31, and 2S). Upon short-pulse excitation with < 600 nm photons, free carries are injected in their respective bands. These carriers thermalize, [3S,4S] recombine with each other (with rate constant of $2.3 \times 10^{-8}$ cm$^3$/s) [24,2S,3S] and descend into shallow (60-100 meV) traps ($\sim 10^{20}$ cm$^{-3}$) with rate constant of $(1-2) \times 10^{-8}$ cm$^3$/s. [24,5S] The scattering time of the plasma is very short, $\approx 0.5$ fs, [31,2S] and no TA signal from the free carriers was observed in the vis and NIR for carrier density < $10^{19}$-$10^{20}$ cm$^{-3}$. [24,2S,3S,5S] For the initial carrier densities of $10^{17}$-$10^{18}$ cm$^{-3}$, the TA signal is dominated by band-tail charges that slowly recombine (with rate constant of $6 \times 10^{-9}$ cm$^3$/s), [24] by hopping and thermal emission, and descend into < $10^{17}$ cm$^{-3}$ of deep traps (such as dangling Si bonds). [24,31,1S,5S] The intraband absorption of these trapped charges is a smooth, featureless curve that gradually ascends from the vis into the NIR. [5S] At low carrier density, the decay kinetics of $\Delta\varepsilon'_\omega(t)$ and $\Delta\varepsilon''_\omega(t)$ are slow (lasting into hundreds of ps) and dispersive. [24,1S,5S] $\Delta\varepsilon'_\omega(t)$ is negative, and the initial phase (for $t<100$ ps) is close to $107^\circ$ (at 1033 nm). [5S]

Fig. 12S shows PPS kinetics obtained for 400 nm excitation of a $d=1.4$ μm thick film of undoped *a*-Si:H. At this excitation energy, the pump light is absorbed in 30 nm near the surface, resulting in high initial density (>$10^{19}$ cm$^{-3}$ for our pump intensities) and considerable excess energy (1.35 eV) of the photocarriers. The center frequency $\omega_0$ of the probe pulse, 800 nm, is matched with the transmission maximum at 796 nm, so the *ΔOD* signal is dominated by photoinduced absorbance. The short-lived "spike" (< 5 ps) has nearly the same decay profile as the (positive) $\Delta\varepsilon'_\omega(t)$ and (negative) $\Delta\varepsilon''_\omega(t)$ in the 310 nm pump - 310 nm probe experiment by Wraback et al. [4S] Similar PPS kinetics with a life time of 1.5 ps for the spike were observed in a 400 nm pump - 2.86 μm probe experiment at Argonne (unpublished). Tauc et al. [31,4S] give an estimate of 2 eV/ps for the rate of carrier relaxation in a-Si:H, which gives 1.5 ps for thermalization time after 400 nm photoexcitation. The decay rate of the "spike" changes with the pump intensity, and the kinetics can be interpreted in terms of a monoexponential process with time constant of ≈2 ps (that Wraback et al. [4S] associate with carrier relaxation) and a bimolecular process with rate constant of $4 \times 10^{-10}$ cm$^3$/s (presumably, due to recombination of these "hot" carriers). For $t$>5 ps, the kinetics show slow decay over > 1 ns; with only a few per cent drop in $\Delta OD$ over the first 50 ps after the 400 nm pulse (i.e., one expects to observe perfectly flat picosecond FDSS kinetics). Such dispersive kinetics are typical for amorphous semiconductors: one can readily find a time window where the kinetics are nearly flat. Thus, for a suitably long pump delay, the wavelength dependence of the function $\mathcal{T}'(\omega)$ (see eq. (A24)) to which $S(\omega)$ (eq. (7)) asymptotically converges, $S(\omega) \approx \mathcal{T}'(\omega)\exp(-\gamma T_e)$ (this formula is justified in Appendix 1), can be found (for this particular system, we have obtained this function at $T$=300 ps). Following other authors, we will assume that $\Delta\varepsilon_\omega$ is constant over the narrow spectral band of the probe pulse, [5S] and the wavelength dependence of $\mathcal{T}'(\omega)$ is due to the $\omega$-dependent Fresnel coefficient $t_\omega$ alone. [1S]

In Figs 13S and 14S(a), FDSS kinetics for the same *a*-Si:H sample are given at several delay times $T$ of the pump pulse and two stretch factors, $s$=-630 ($\tau_{GVD} = 0.53\,fs$) and $s$=-3,780 ($\tau_{GVD} = 1.23\,fs$). In Fig. 13S(b), $T$=0 ps and $T$=9 ps kinetics obtained for $s$=-630 (shown in Fig. 13S(a)) were normalized by the $T$=300 ps kinetics [that yield the spectral profile of $\mathcal{T}'(\omega)$]. These normalized kinetics are flat after the first few ps, suggesting that the normalization procedure succeeds in compensating for the curved transmission profile. In Fig. 14(b), trace (i) the two oscillation patterns are juxtaposed in time (given in the units of $\tau_{GVD}$). In the frequency domain, $T$=0 ps corresponds to the time origin placed at the transmittance maximum whereas $T$=9 ps corresponds to this origin placed at the reflectance maximum (Fig. 13S(b)). Despite a considerable change in the phase of the complex factor $t_\omega^{-1}(\partial t_\omega / \partial \varepsilon)$ for these two positions, the two oscillation patterns are almost exactly the same, validating the theoretical analysis given at the end of Appendix 1. The same applies to the kinetic traces obtained for a greater compression factor (Fig. 14S(a) and Fig. 14S(b), trace (ii)).

We conclude that for thin-film samples that exhibit closely spaced interference fringes (comparable to the spectral width of a femtosecond probe pulse), meaningful "TA kinetics" can still be obtained from the FDSS traces simply by doing the $S(\omega)/\mathcal{T}'(\omega)$ normalization. These reconstructed kinetics are unique in the sense that their oscillation patterns and overall shape do not change with the position of the probe band with respect to the fringe pattern. Like the PPS kinetic traces in such a situation, these "TA kinetics" are a combination of transient absorption and reflection kinetics averaged over the probe band. Unlike PPS, FDSS simultaneously yields the spectrum of $\mathcal{T}'(\omega)$ needed to obtain phase relations that are required for the kinetic analysis.

## 3. Additional references

## 4. Figure captions (1S to 14S)

Fig. 1S

FDSS kinetics $S(\omega)$ given by eq. (A17) for a thin-film sample with $\eta=3.44$, $\kappa=1.1 \times 10^{-3}$,[24] and $d=1.27$ μm, for three pump delays: $T=0$ ps (i), 20 ps (ii), and 35 ps (iii). Traces (a) and (b) are the functions $\mathcal{T}(\omega)$ given by eq. (A24) for pure photoabsorption, $\phi_\varepsilon=90°$ (a), and photorefraction, $\phi_\varepsilon=0°$ (b). Other simulation parameters were $\delta=0$ (infinite spectral resolution), $\Delta\varepsilon''_\omega \equiv 1, \tau_p = 20\,fs,\ \tau_L = 100\,fs,\ s = 2,048,$ and $\gamma^{-1} = 40\,ps$.

Fig. 2S

(a) Same as Fig. 1S, but FDSS kinetics were normalized by $\mathcal{T}(\omega)$. For $\gamma T_e \gg 1$, these normalized kinetics asymptotically approach $\exp(-\gamma T_e)$. (b) A comparison between the normalized oscillation patterns.

Fig. 3S.

Center wavelength $(\lambda)$ dependence of (a) the Littrow angle (dotted lines) and the beam spread (thick solid lines) on the grating and (b) stretch factor (thick solid lines; negative for a compressor) and TOD (dotted lines) for $\tau_p$=20 fs probe pulse (±267 cm$^{-1}$ "optimum range") dispersed using a 1200 g/mm grating in the $m$-th diffraction order at $L_g$=100 cm (the GVD and the beam spread are proportional to this length). The TOD is conveniently given as $-\xi_3/(\omega_0\tau_p)$, where $\omega_0 = 2\pi\lambda/c$.

Fig. 4S.

(a) *Filled circles:* Pump-probe kinetics of TA obtained for 400 nm excitation of a 1mm thick polycrystalline ZnSe sample (detected with a transform limited 30 fs FWHM, 800 nm probe pulse). The line drawn through the points is a biexponential fit (see the text). The fast (14 ps) component corresponds to the free carrier trapping; the slow (910 ps) component corresponds to the recombination of trapped charges. Nearly 80% of this "absorbance" signal is due to the negative change in the refraction index due to the free carriers. (b) A family of 512-channel FDSS kinetics obtained from the same sample under identical excitation conditions. In this particular run, the compression factor is 2,048 and $\tau_{GVD}$ is 905 fs (so that 1 ps is equivalent to 6 cm$^{-1}$); the whole kinetics covers 530 cm$^{-1}$. Each FDSS kinetics is the average of 150K shots. Filled circles indicate the $T=30$ ps trace that was used to determine the spectral response of the sample; the solid line drawn through the signal was polynomially smoothed and used to normalize other FDSS kinetics. The resulting normalized kinetic traces are shown in Fig. 9. The premise of this normalization procedure is that the $T=30$ ps kinetics are flat; this approximation is justified by the PPS kinetics given above. Similar normalization was done for other systems too.

Fig. 5S.

(a) Pump-probe TA of hydrated electrons observed at the delay time of 7 ps (filled circles) and 400 ps (empty squares) upon 400 nm photoexcitation of 75 mM aqueous KI in a 150 µm optical path high-speed jet (double logarithmical plot). The beam radii of the 400 nm ($\tau_L$=200 fs) pump and 800 nm ($\tau_p$=20 fs) probe were 35 and 19.5 µm, respectively. Upon the photoexcitation, iodide donates the electron to water; the resulting electron fully thermalizes and localizes within 2 ps (so that the 7 ps measurement gives the initial yield of thermalized electrons). The geminate pair of the electron and the residual iodine atom slowly recombine on the sub-ns time scale; the 400 ps absorbance gives an estimate for the escape yield of these hydrated electrons. For pump irradiance < 0.3 TW/cm$^2$, both absorbance signals increase as the second power of the irradiance (solid lines with the slopes of 2±0.04 and 1.94±0.04 for 7 and 400 ps, respectively), as expected for a biphotonic process; the fraction of electrons that escape the geminate recombination does not change with the pump irradiance. For pump irradiance > 1 TW/cm$^2$, the electron yield increases linearly with the irradiance. As the irradiance increases, the absorbance signals at 7 and 400 ps first get closer, indicating less efficient geminate recombination, then diverge again, signaling the onset of cross-recombination (*the inset*). The change in the kinetics is illustrated in (b) for traces (i) *(to the left)* and (ii) *(to the right)* that were obtained at 0.22 and 1.91 TW/cm$^2$, respectively. The vertical bars give 95% confidence limits; the solid lines are biexponential fits. Slower geminate recombination at the higher pump irradiance is indicative of the "2+1" photoprocess in which the extra 400 nm photon is absorbed by pre-thermalized electron that is thereby excited deep into the conduction band and localized away from the iodine atom (see the text).

Fig. 6S.

(a) A PPS kinetics obtained for 400 nm excitation of 75 mM aqueous iodide flowing in a 5 mm optical path cell. The TA signal is detected using an 800 nm, 33 fs (*trace (i), filled squares*) and 4 ps (*trace (ii), filled circles*) FWHM probe. The pump irradiance was < 0.1 TW/cm$^2$, i.e., these kinetics were obtained in the 2-photon excitation regime (see the legend to Fig. 5S). (b) A family of FDSS kinetics obtained for the same system ($\tau_{GVD}$= 890 fs, 1 ps = 5.6 cm$^{-1}$). The delay times $T$ of the pump are given in the legend; the kinetics were normalized by the $T$=260 ps trace to correct for the nonflat spectral response. Arrows indicate the "spike" on the rising edge of the FDSS kinetics due to the rapid swing of the phase of dielectric function in the course of electron thermalization (see Fig. 10 and 9S and the discussion in section V.2).

Fig. 7S.

(a) The data of Fig. 6S(b) replotted on an "absolute" time scale (with the "delay time" given by $T_e$-$T$); the spliced FDSS kinetics faithfully reproduce the PPS kinetics replotted from Fig. 6S(a). The vertical bars indicate 95% confidence limits for the PPS trace. (b) $T$=0 ps and $T$=20 ps traces replotted from Fig. 6S(b) vs. the "absolute" time.

Fig. 8S.

Normalized FDSS kinetics obtained from 75 mM aqueous KI flowing in a 150 μm jet; similar excitation conditions to those specified in Fig. 5S(b). The 400 nm pump irradiance (TW/cm$^2$) and the maximum $\Delta OD$ at 800 nm were, respectively (i) 0.72 and 0.043, (ii) 1.4 and 0.147, and (iii) 2.43 and 0.3. These kinetics were obtained for GVD of −1.58 ps$^2$ ($\tau_{GVD}$=1.26 ps; 1 ps = 3.36 cm$^{-1}$); 75K pump on - pump off shots were averaged for each kinetic trace. As the irradiance increases, the decay kinetics of the hydrated electron become flatter (compare with the PPS kinetics given in Fig. 5S(b)) and the amplitude of the oscillations becomes smaller.

Fig. 9S.

Normalized FDSS kinetic traces shown in Fig. 10 replotted as a function of the reduced time $(T_e - T)/\tau_{GVD}$. Traces (i) to (iv) correspond to traces (a) to (d) in Fig. 10, respectively. Vertical ticks in traces (ii) and (iv) indicate 95% confidence limits for the FDSS kinetics. Arrows indicate the position of the "spike" (see the text).

Fig. 10S.

(a) A typical power dependence of the hydrated electron yield in 400 nm photoionization of neat liquid water (150 μm thick jet) observed via the PPS-detected electron absorbance at 800 nm, ca. 16 ps after the 200 fs FWHM excitation pulse (at which time the thermalization is complete). The pump and probe beam radii are 56 and 14 μm, respectively. At low irradiance (< 0.5 TW/cm$^2$), the ionization is 3-photon and results in the geminate decay kinetics for which the escape yield of the electron is ca. 72%. At higher irradiance, the yield linearly increases with the pump power, indicating the occurrence of the "3+1" excitation process. Simultaneously, the time profile of the kinetics changes so that the escape yield approaches > 90%. (b) TA kinetics obtained in the "3+1" regime (for 80 μJ pump pulse). The decay is second order *(solid line)* and originates mainly through the cross-recombination of the electrons and hydroxyl radicals (that occurs with a rate constant of 3x10$^{10}$ M$^{-1}$ s$^{-1}$). The path-average concentration of the electrons is 1.6 mM, which gives a time constant of 21 ns; the observed time constant is higher, ca. 5 ns, due to the extremely non-homogeneous excitation profile for the "3+1" photoprocess. The vertical bars are 95% confidence limits.

Fig. 11S.

(a) Normalized FDSS kinetics obtained for 3 x 400 nm photon excitation of N$_2$-saturated liquid water flowing in a high-speed jet (solid line). Filled circles indicate the pump probe kinetics obtained under the same excitation conditions (vertical bars indicate 95% confidence limits). The maximum $\Delta OD$ is ca. 1x10$^{-2}$; see the text for other parameters. An arrow indicates the "spike" that is quite analogous to the same "spike" in the FDSS kinetics obtained for iodide CTTS (e.g., Fig. 6S). (b) Power dependence for normalized FDSS kinetics obtained under the same "3+1" excitation conditions as the PPS kinetics shown in Fig. 10S(b). The pump power and the maximum $\Delta OD$ were (i) 27 μJ and 0.017, (ii) 49 μJ and 0.2, and (iii) 94 μJ and 0.394, respectively. Note the drastic reduction in the oscillation amplitude at the higher optical density. The FDSS kinetics were obtained for $\tau_{GVD}$=1.21 ps (1 ps = 3.6 cm$^{-1}$) with a compressed 33 fs FWHM probe pulse. Traces (i), (ii), and (iii) are the averages of 30K, 45K, and 100K shots, respectively.

Fig. 12S.

Pump-probe kinetics observed upon the 400 nm excitation of 1.4 µm film of amorphous hydrogenated silicon on a suprasil substrate (801.3 nm detection). The pump power was (i) 9 and (ii) 43 µJ. After the initial rapid decay (in 3 ps), a slower decay due to the recombination of trapped charges in the bulk is observed. The 33 fs FWHM probe pulse is centered at the transmission extremum (see Fig. 13S) and has a band pass similar to the fringe spacing. Note the flatness of the decay kinetics for t>10 ps at the lower excitation power.

Fig. 13S.

(a) FDSS "kinetics" for the *a*-Si:H system (see the previous caption) plotted as a function of $\Delta\omega$. These kinetics were obtained with a 400 nm pulse of 19 µJ and *T*=0 (ii) and +9 ps (i), respectively. The stretch factor for a 33 fs FWHM probe pulse is -627, $\tau_{GVD}$=528 fs (1 ps = 19 cm$^{-1}$). The spectral response (iii) was obtained from a *T*=300 ps trace. (b) The kinetics from Fig. 13S(a) normalized by this spectral response (shown to the right and to the top).

Fig. 14S.

(a) Same as Fig. 13(b), for a different spot on the same sample and same excitation conditions. FDSS traces (i) and (ii) were obtained for *T*=0 (ii) and +50 ps (i), respectively. The maximum $\Delta OD$ is ca. 6x10$^{-2}$. The kinetics were obtained for the stretch factor of –3782 and $\tau_{GVD}$=1.23 ps (1 ps = 3.6 cm$^{-1}$). (b) FDSS kinetics at *T*=0 ps *[lines and symbols]* and T= 9 ps (i) and 50 ps (ii) *[lines only]* in Figs. 13S(b) and 14S(a) replotted vs. the reduced time, $(T_e - T)/\tau_{GVD}$. The stretch factors are given in the figure. $\tau_{GVD}$ is (i) 0.53 and (ii) 1.23 ps, respectively. Compare this figure with Fig. 2S.

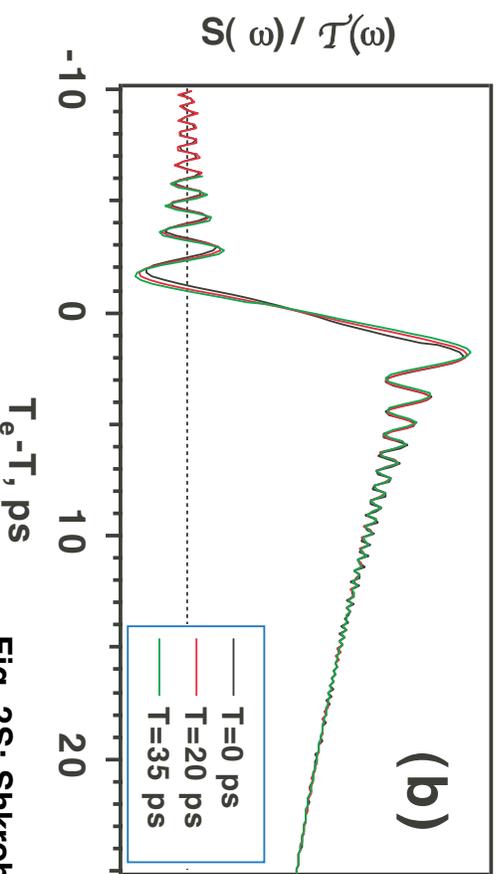

Fig. 1S; Shkrob et al

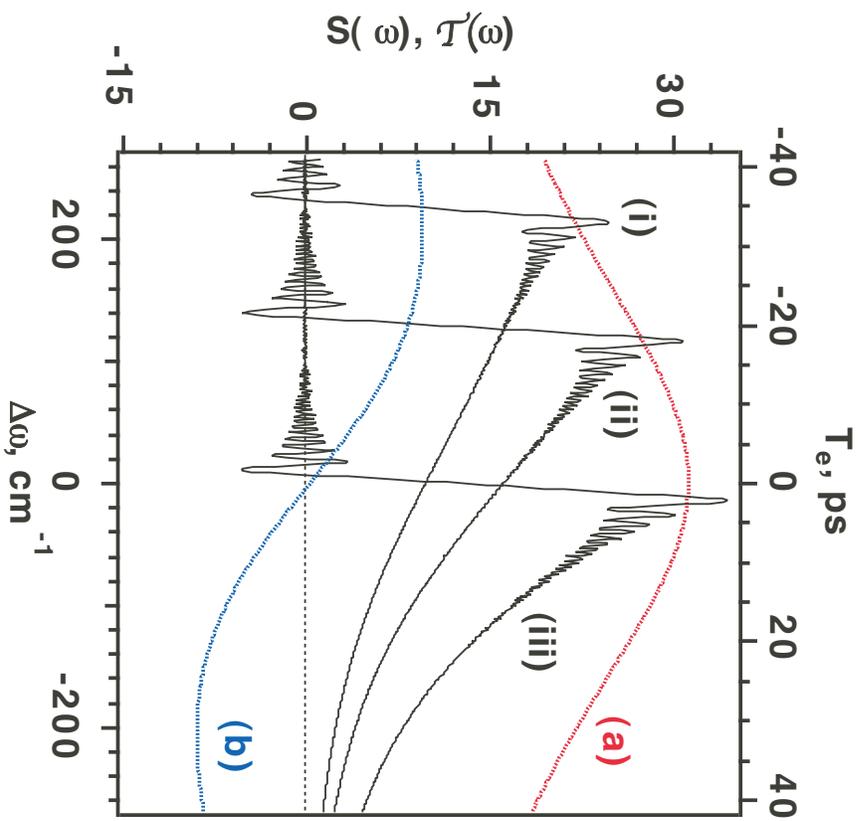

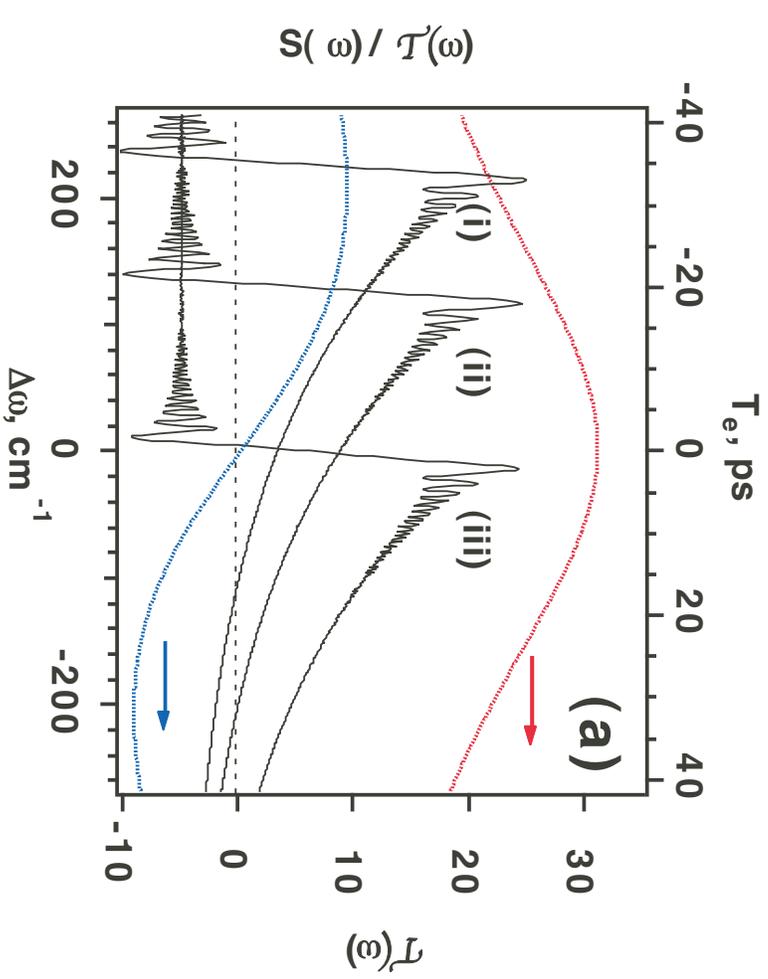

Fig. 2S; Shkrob et al

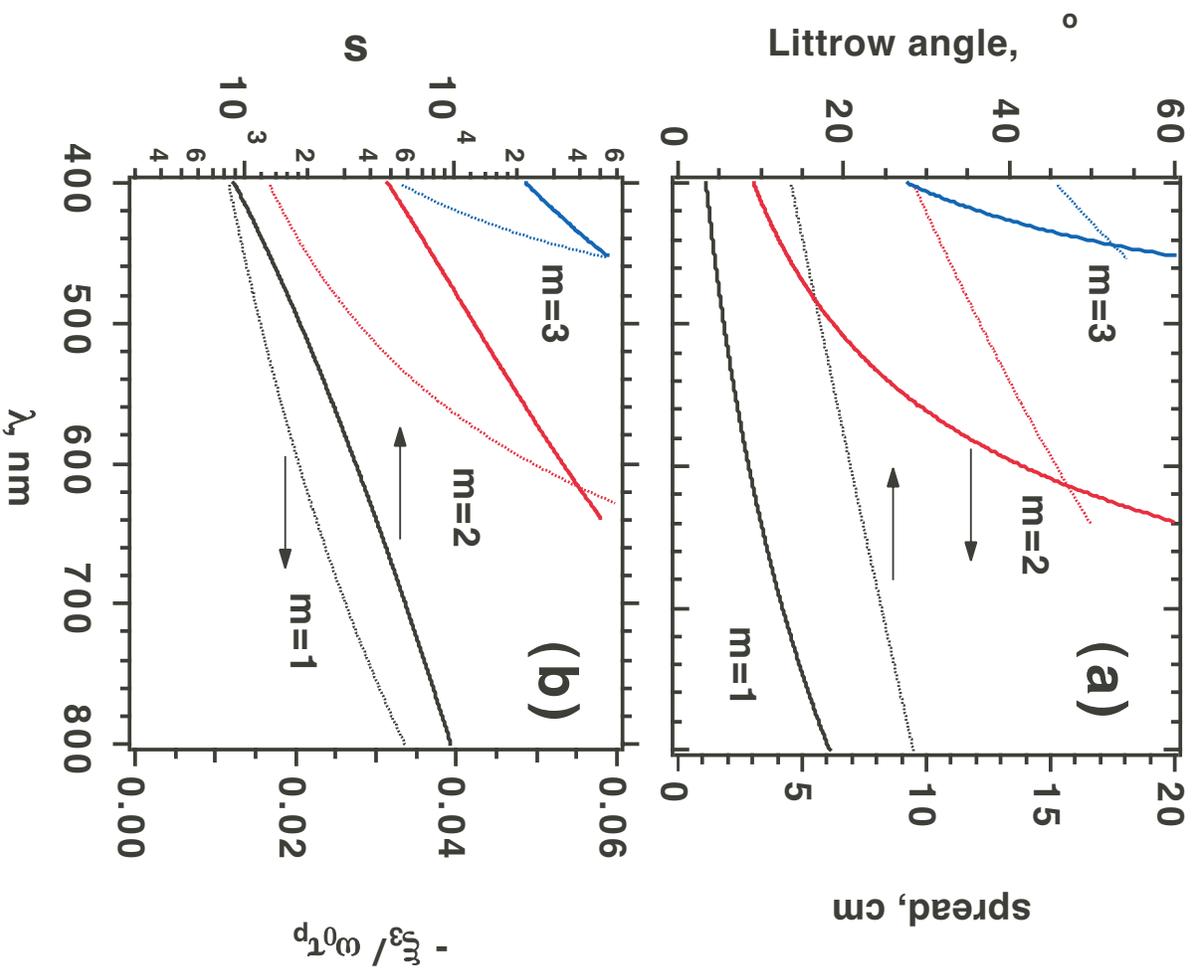

Fig. 3S; Shkrob et al

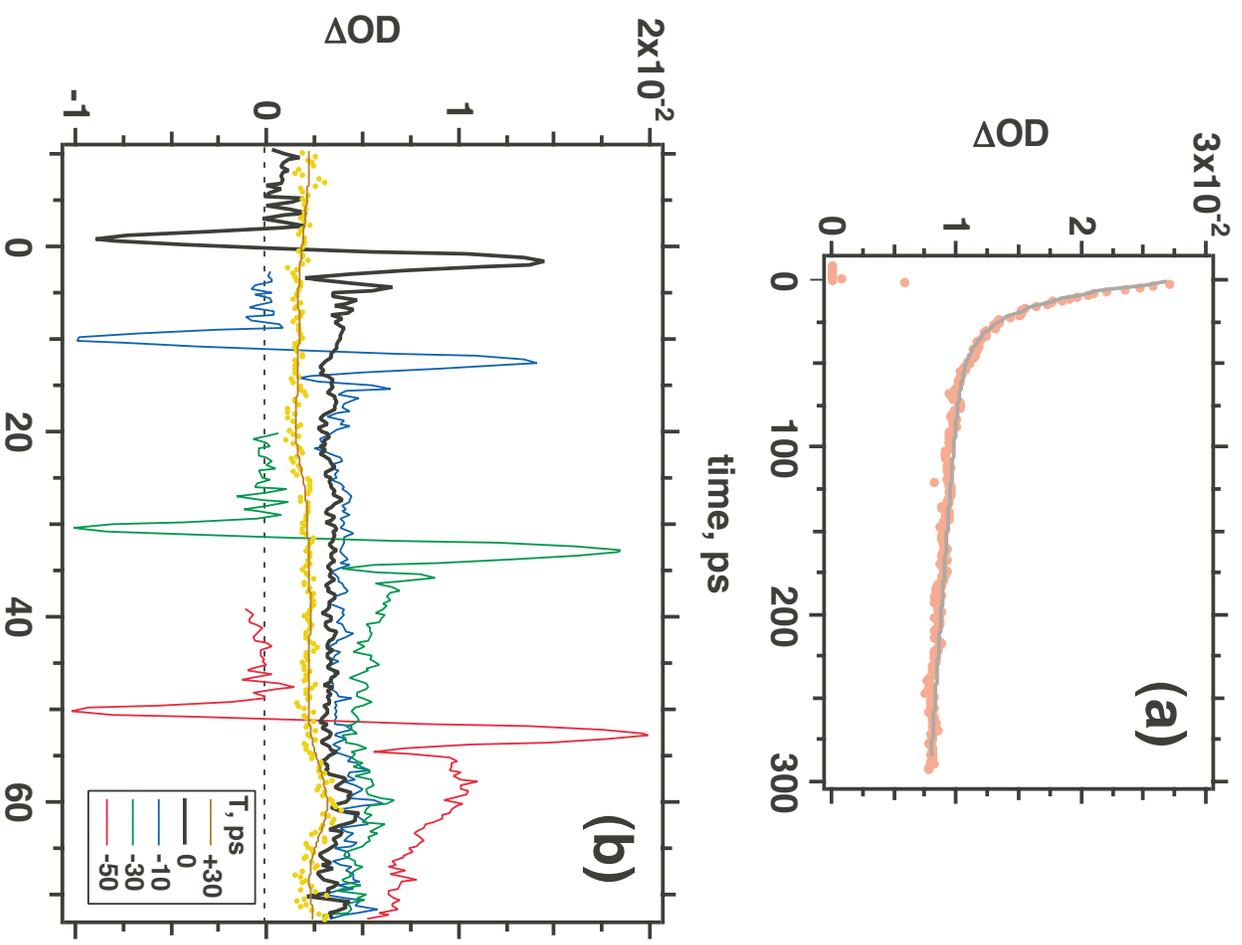

Fig. 4S; Shkrob et al

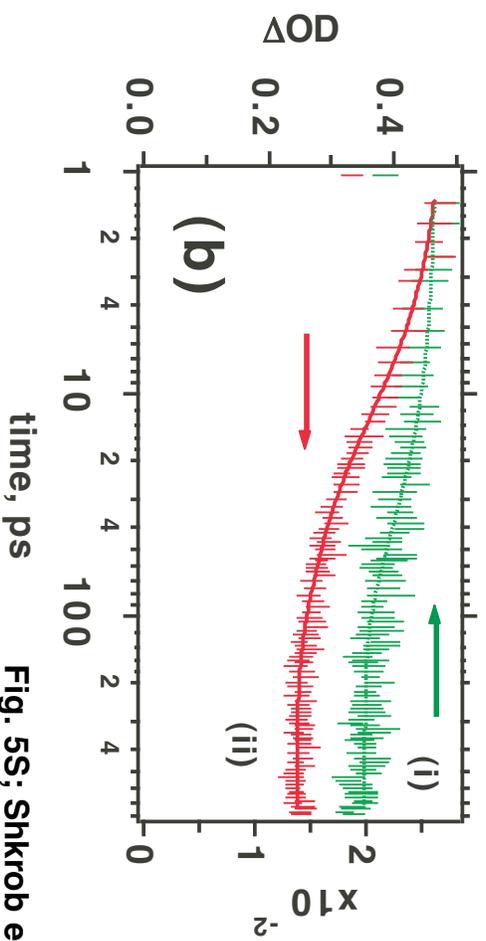

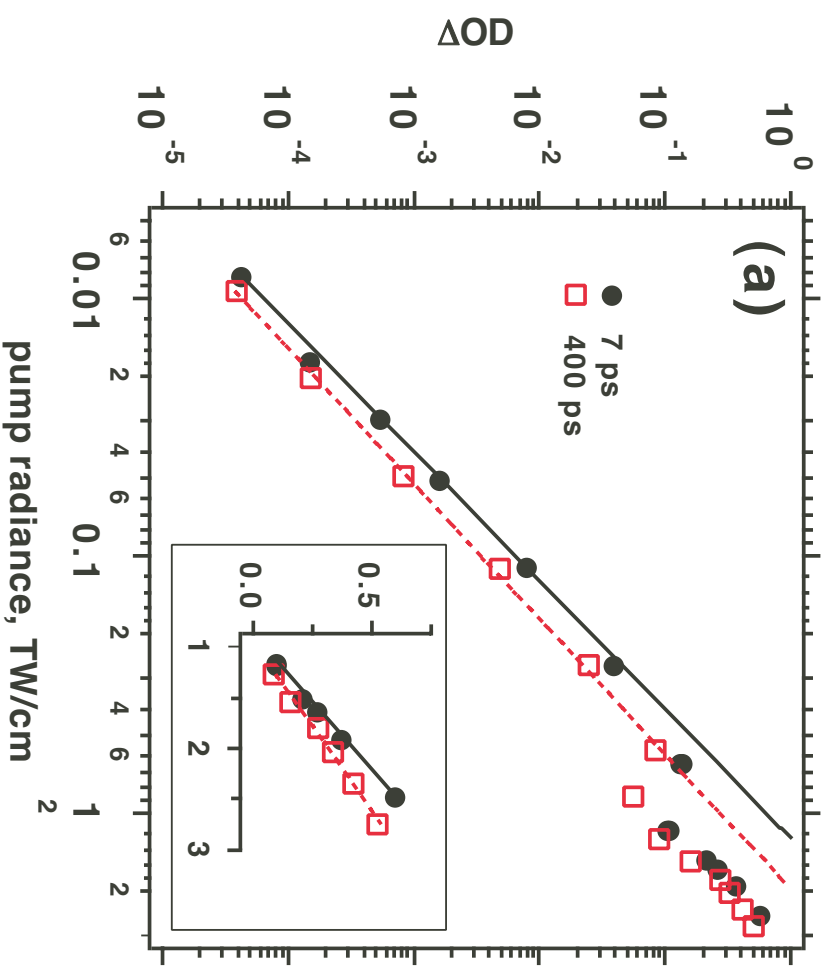

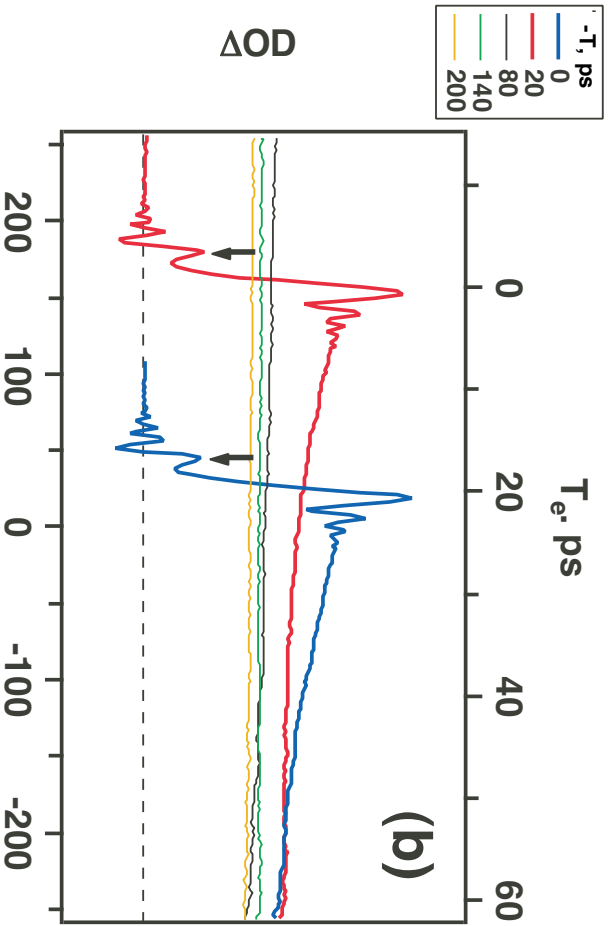

Fig. 5S; Shkrob et al

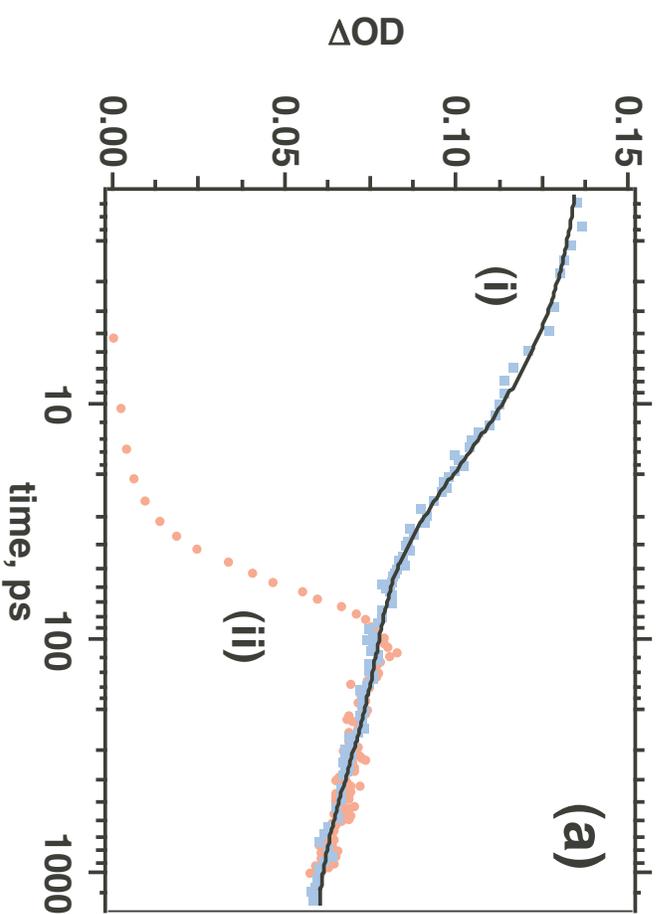

Fig. 6S; Shkrob et al

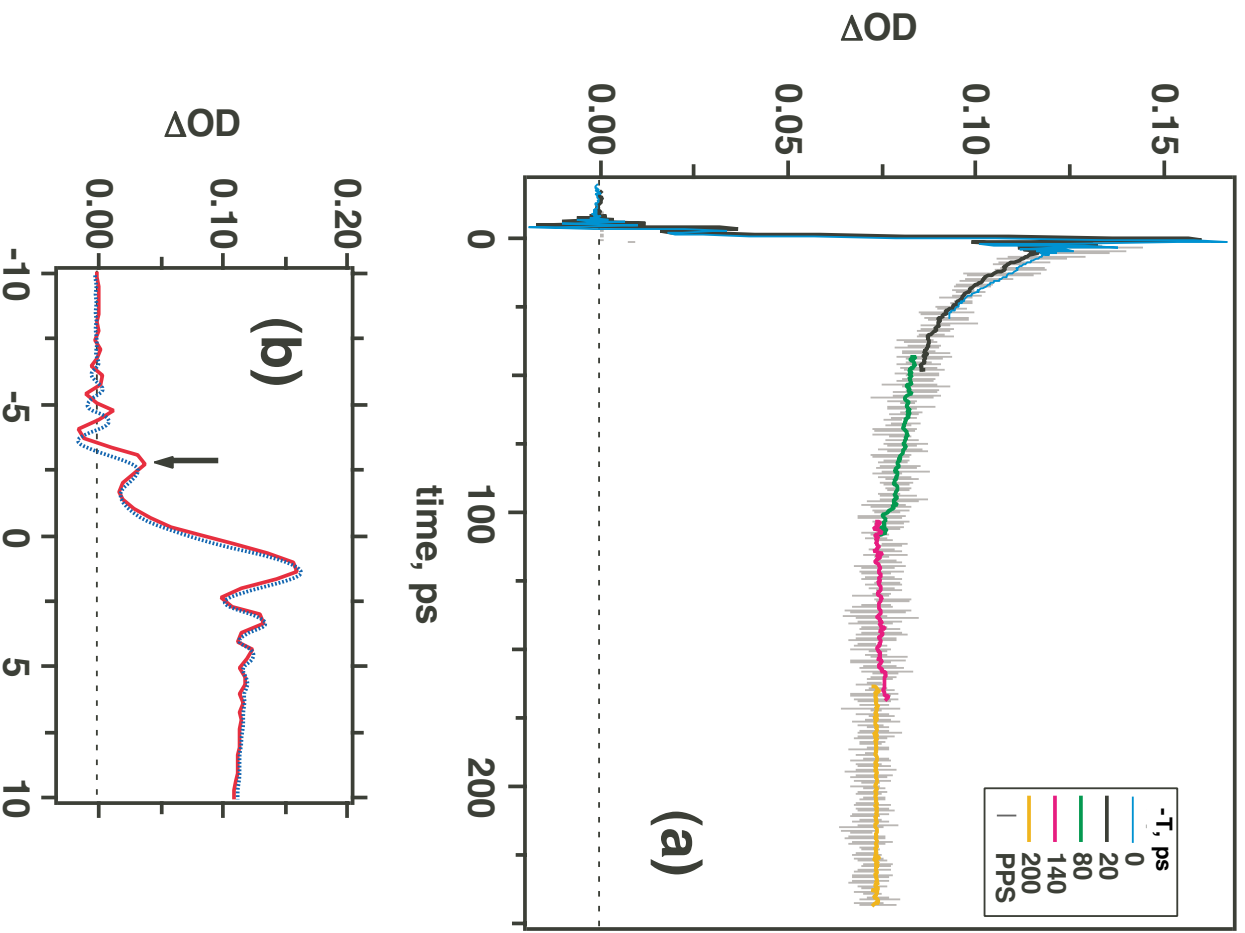

Fig. 7S; Shkrob et al

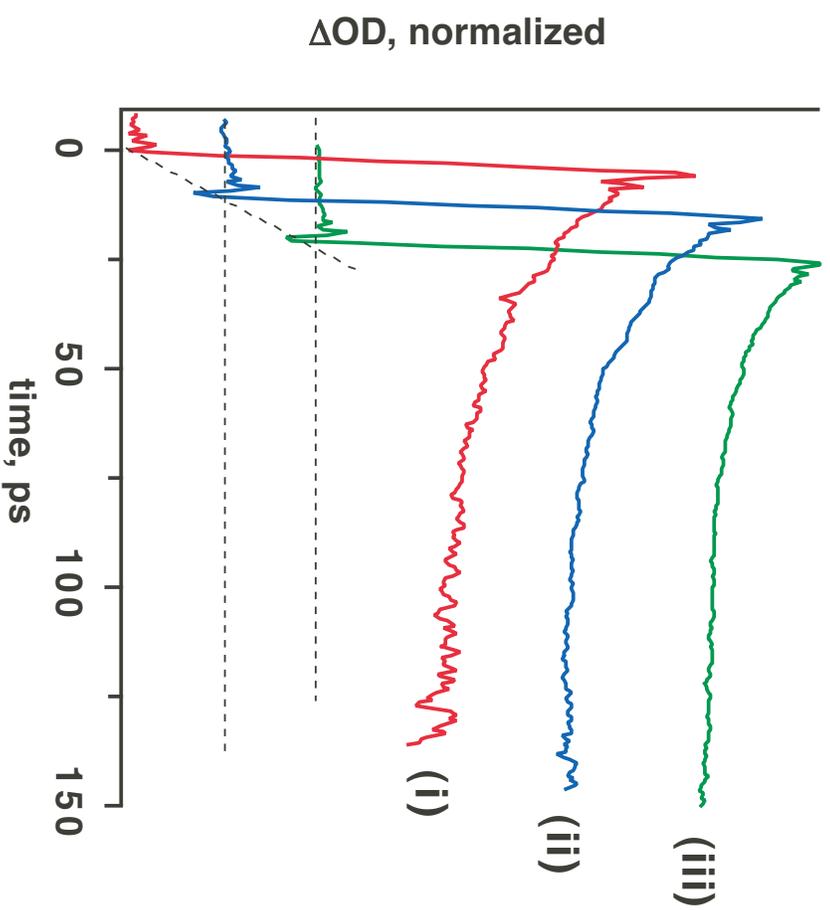

Fig. 8S; Shkrob et al

Fig. 9S; Shkrob et al

Fig. 10S; Shkrob et al

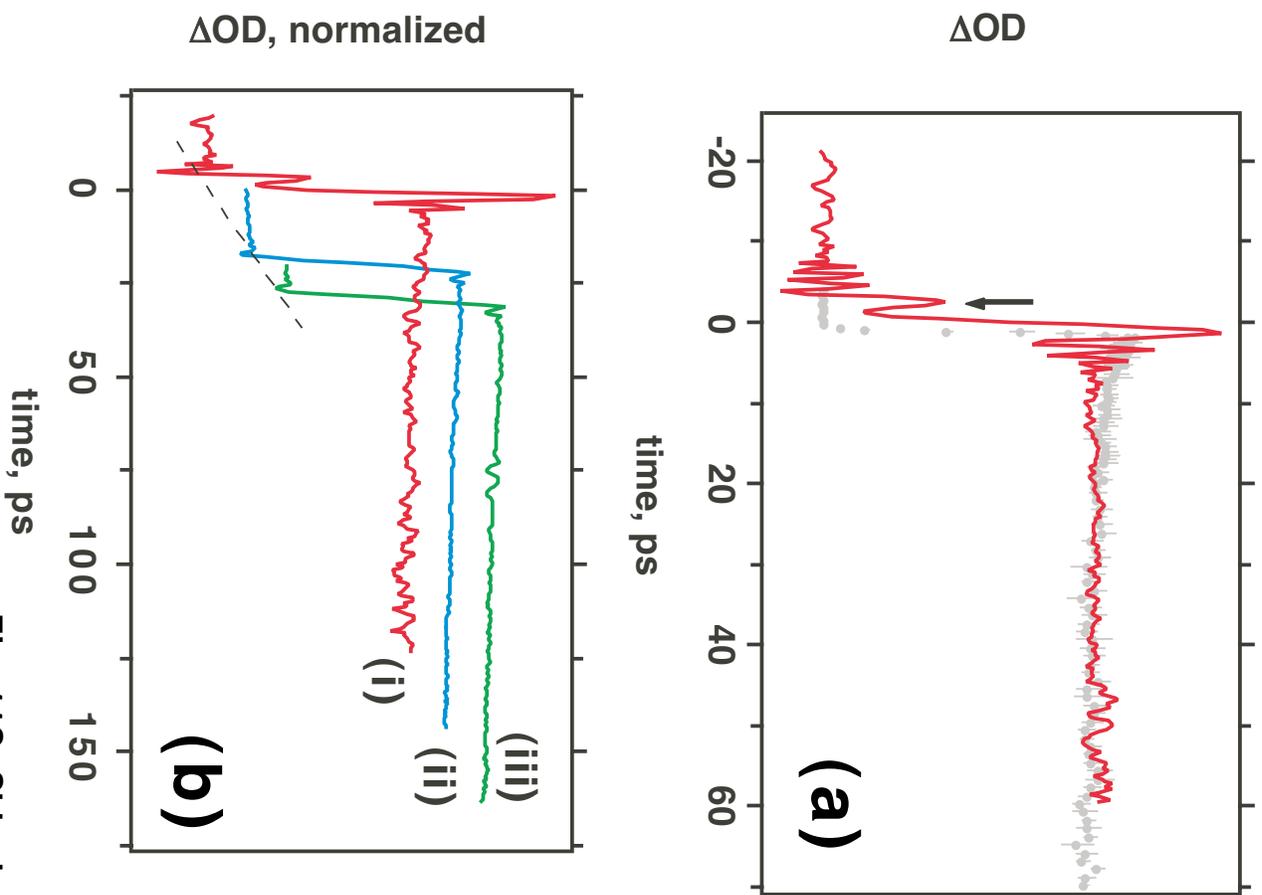

Fig. 11S; Shkrob et al

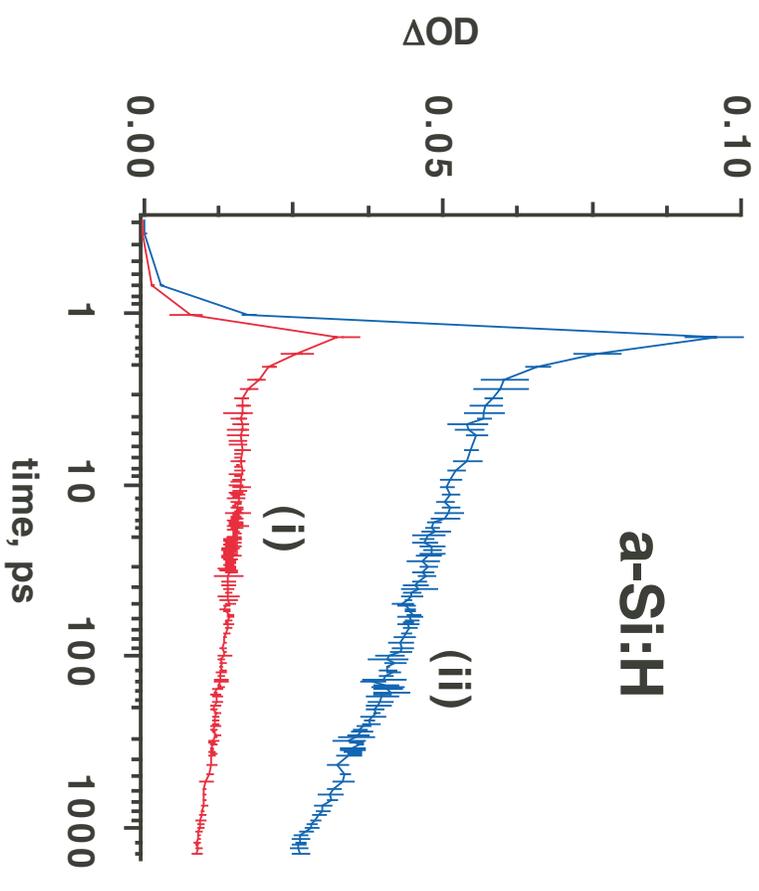

Fig. 12S; Shkrob et al

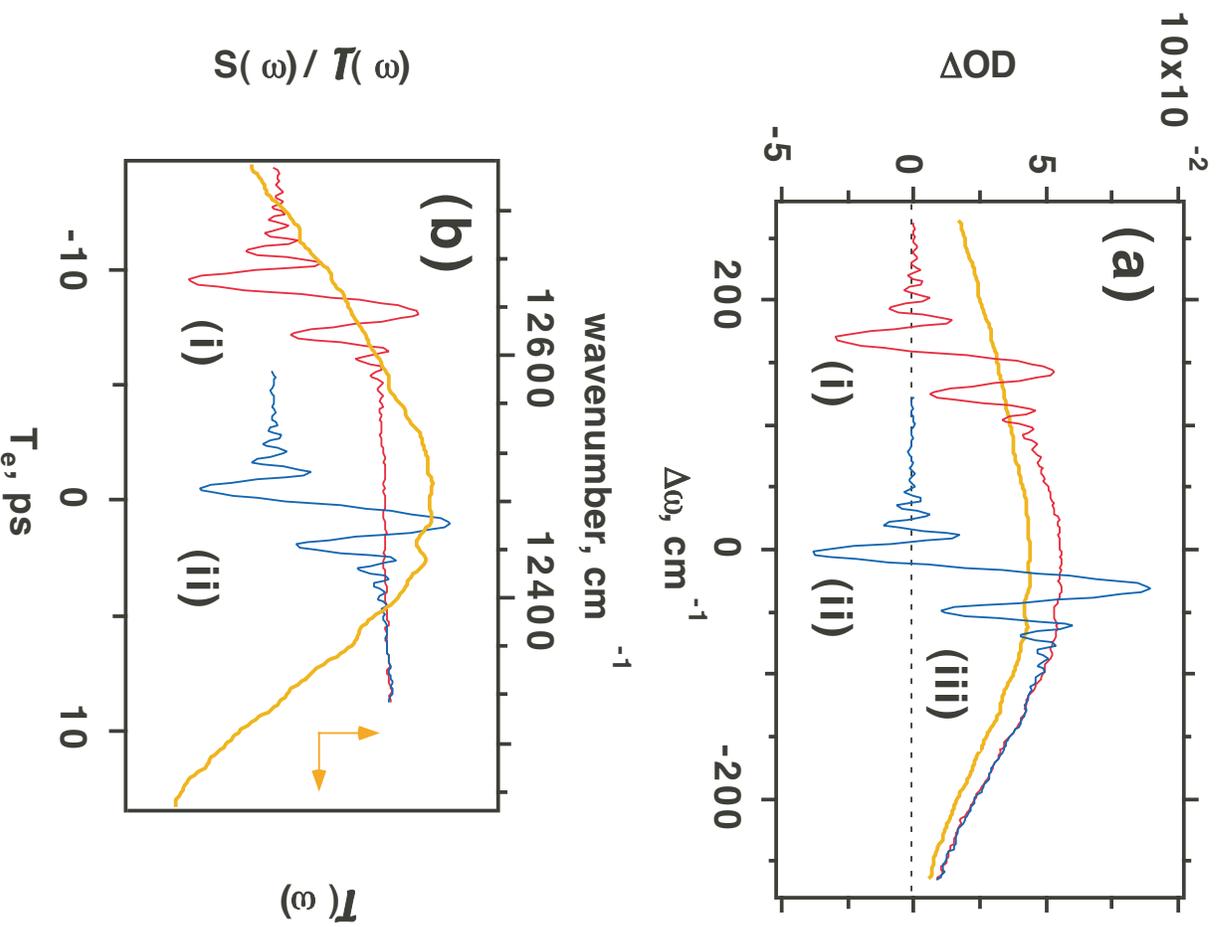

Fig. 13S; Shkrob et al

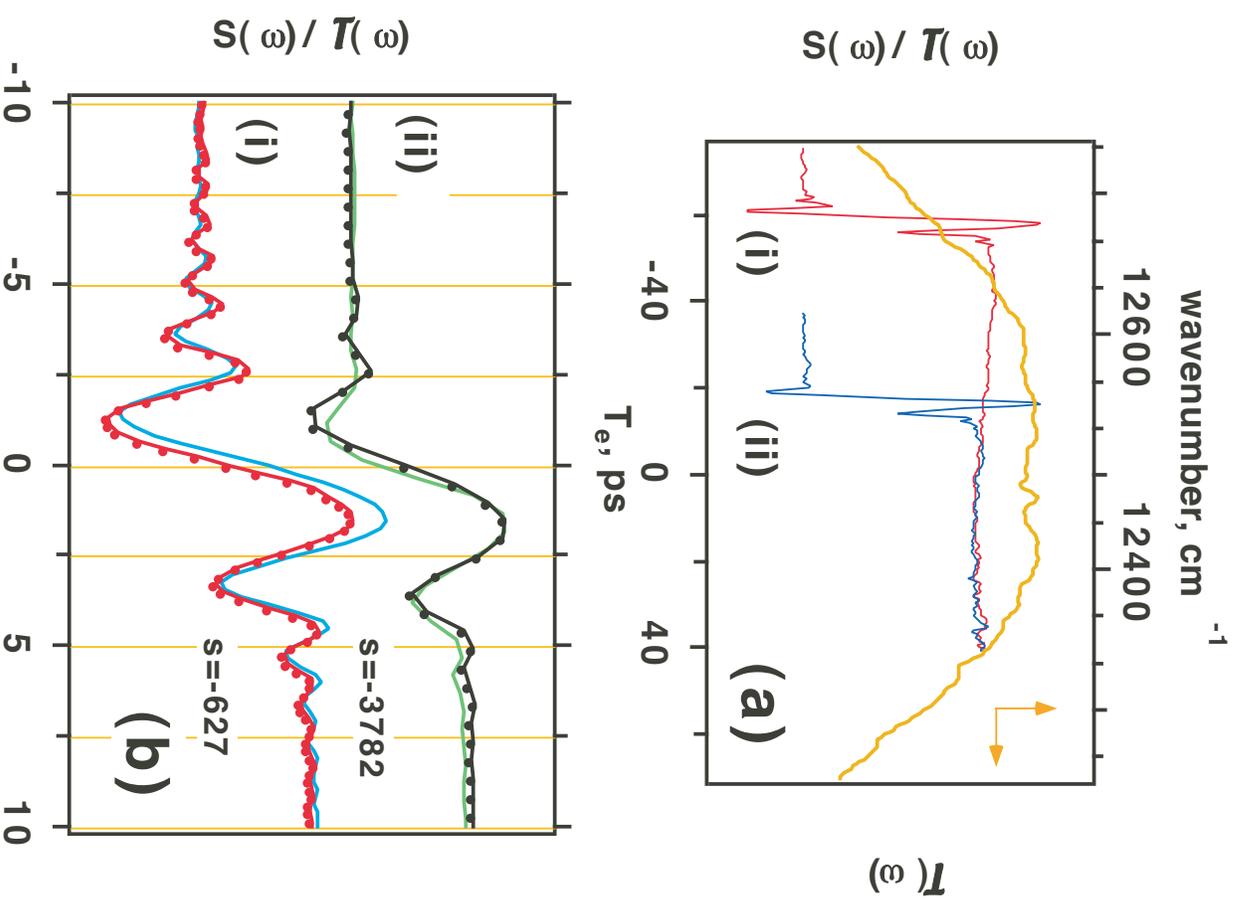

Fig. 14S; Shkrob et al